\documentclass[fleqn,usenatbib]{mnras}
\DeclareSymbolFont{cmletters}{OML}{cmm}{m}{it}
\DeclareMathSymbol{v}{\mathalpha}{cmletters}{"76}

\usepackage{ae,aecompl}
\usepackage{times}
\usepackage{soul}

\usepackage{tabularx,ragged2e,booktabs,caption}
\usepackage{amsmath}
\usepackage{amssymb}
\usepackage{epsfig}
\usepackage{graphicx}
\usepackage{ifthen}
\usepackage{latexsym}
\usepackage{rotating}
\usepackage{times,epsf}
\usepackage{txfonts}
\usepackage{varioref}
\usepackage{verbatim}
\usepackage{array}
\usepackage{url}
\usepackage{color}
\usepackage[T1]{fontenc}
\usepackage{epstopdf}
\usepackage{ulem}

\def\gsim{\mathrel{\raise.5ex\hbox{$>$}\mkern-14mu
             \lower0.6ex\hbox{$\,\sim$}}}
\def\lsim{\mathrel{\raise.3ex\hbox{$<$}\mkern-14mu
             \lower0.6ex\hbox{$\,\sim$}}}

\newcommand{\be}{\begin{equation}}
\newcommand{\ee}{\end{equation}}
\newcommand{\bea}{\begin{eqnarray}}
\newcommand{\eea}{\end{eqnarray}}

\title{Hybrid Numerical Simulations of Pulsar Magnetospheres}
\author[I. Contopoulos, J. P\'{e}tri, P. Stefanou]
       {I. Contopoulos$^{1,2}$\thanks{E-mail: icontop@academyofathens.gr}, J. P\'{e}tri$^3$, P. Stefanou$^4$\\
$^1$ Research Center for Astronomy and Applied Mathematics, Academy of Athens, Athens 11527, Greece\\
$^2$ National Research Nuclear University (MEPhI), Moscow 115409, Russia\\
$^3$ 
Observatoire astronomique de Strasbourg, Universit\'{e}  de Strasbourg, CNRS, UMR 7550, 11 rue de l’ universit\'{e}, F-67000 Strasbourg, France\\
$^4$ Section of Astrophysics, Astronomy and Mechanics, Department of Physics, University of Athens, Athens 15783, Greece
}

\begin{document}

\maketitle

\label{firstpage}

\begin{abstract}
We continue our investigation of particle acceleration in the pulsar equatorial current sheet (ECS) that began with \cite{C19} and \cite{CS19}. Our basic premise has been that the charge carriers in the current sheet originate in the polar caps as electron-positron pairs, and are carried along field lines that enter the equatorial current sheet beyond the magnetospheric Y-point. In this work we investigate further the charge replenishment of the ECS. We discovered that the flow of pairs from the rims of the polar caps cannot supply both the electric charge and the electric current of the ECS. The ECS must contain an extra amount of positronic (or electronic depending on orientation) electric current that originates in the stellar surface and flows outwards along the separatrices. We develop an iterative hybrid approach that self-consistently combines ideal force-free electrodynamics in the bulk of the magnetosphere with particle acceleration along the ECS. We derive analytic approximations for the orbits of the particles, and obtain the structure of the pulsar magnetosphere for various values of the pair-formation multiplicity parameter $\kappa$. For realistic values $\kappa\gg 1$, the magnetosphere is practically indistinguishable from the ideal force-free one, and therefore, the calculation of the spectrum of high-energy radiation must be based on analytic approximations for the distribution of the accelerating electric field in the ECS.
\end{abstract}


\section{Introduction}

We continue our investigation of electromagnetic (Poynting) energy dissipation in the axisymmetric pulsar magnetosphere following the `hybrid' approach of \cite{C07a,C07b,CKK14}, \cite{C19} (hereafter Paper~I), and \cite{CS19} (hereafter Paper~II). The pulsar magnetosphere is considered to be everywhere ideal and force-free {\it except} in a dissipative layer that develops beyond the tip of the closed line region along the equatorial current sheet (hereafter ECS). The ECS is threaded by magnetic field lines that originate around the rim of the polar cap and contain a finite amount of magnetic flux
\begin{eqnarray}
&&\Psi_{\rm ECS}=2\pi r_{\rm pc}\delta B_*=\frac{2\delta}{r_{\rm pc}}\Psi_{\rm open}\ll \Psi_{\rm open}=\pi r_{\rm pc}^2 B_*\ .
\label{rim}
\end{eqnarray}
Here, $\Psi_{\rm open}\approx 1.25 \Psi_{{\rm open}\ {\rm dipole}}$ is the total amount of magnetic flux that crosses the light cylinder at a distance $r_{\rm lc}\equiv c/\Omega$, $\Psi_{{\rm open}\ {\rm dipole}}\equiv \pi r_*^3 B_*/r_{\rm lc}$ is the amount of dipole magnetic flux that crosses the equator beyond the light cylinder  \citep{CKF99,S06,T06}, $r_{\rm pc}\approx \sqrt{1.25} r_{{\rm pc}\ {\rm dipole}}\equiv \sqrt{1.25} r_*^3/r_{\rm lc}$ is the radius of the so-called `polar cap', and $\Omega$ is the angular velocity of stellar rotation.
These magnetic field lines carry the electrons and positrons required to support the electric current of the ECS, and transfer electromagnetic energy from the central `generator' (the stellar rotation) to the electrons and positrons in the ECS. The thickness $\delta$ of the polar cap rim that supplies the ECS with charge carriers and electromagnetic energy is inversely proportional to the pair-formation multiplicity $\kappa\gg 1$ (how many pairs are produced per Goldreich-Julian charge particle in the polar cap; Papers~I \& II). Without loss of generality, we will only consider aligned rotators with $B$ along $\Omega$ at the poles.

In our `hybrid' approach, particle orbits are only considered in the dissipative ECS where positrons are accelerated outwards and electrons inwards. Electrons and positrons are in general extremely relativistic (Lorentz factors $\gg 10^3$), and, during their acceleration by the radial electric field that develops in the ECS, they both radiate high-energy radiation along the direction of their motion. There is no point to follow their motion in the rest of the ideal magnetosphere where they simply flow along magnetic field lines and drift across them with drift velocity $c{\bf E}\times {\bf B}/B^2$ and gyroradii much smaller than the macroscopic dimensions of the magnetosphere\footnote{Equivalently, this drift is the definition of field line velocity and dragging of particles by the magnetic field.}. The main reason we opted for this `hybrid' approach (ideal force-free everywhere with consideration of particle dynamics only in the ECS) is that we believe it is too early for an ab initio reconstruction of the pulsar magnetosphere with PIC numerical simulations \citep{C16}. This is due to insufficient numerical resolution \citep[a few hundred grid points inside the light cylinder is grossly inadequate as has been shown clearly in figure~1 of][]{TSL13} and unphysical simulation parameters (Larmor radii on the order of the light cylinder radius instead of at least nine orders of magnitude smaller, Lorentz factors smaller than about $10^3$ instead of at least five orders of magnitude larger, etc.). Moreover, it is not clear whether the dissipation obtained with present day numerical PIC codes 
($\sim 30\%$ of $\dot{E}$ within a few $r_{\rm lc}$ just outside the light cylinder) is indeed physical \citep[as e.g. in][]{CKK14}, or numerical (compare e.g. fig.~6 of \cite{CPPS15} with fig.~13 of \cite{PBH12} and fig.~1c of \cite{TSL13}).
This makes them inadequate to study the physical electromagnetic energy dissipation without a deeper understanding of the physical processes that take place in that region.

In the present work we will improve the `ring-of-fire' model proposed in Paper~II. In that model, we had assumed for simplicity that the dissipation layer (denoted by DL in that paper) had a finite radial extent at the origin of the ECS beyond the tip of the closed-line region near the light cylinder. Beyond that region, the ECS was considered dissipationless all the way to infinity. We will now relax that assumption since it seems more natural that the ECS is everywhere dissipative.
\begin{figure}
 \centering
 \vspace{-0.32cm}
 \includegraphics[width=8cm,height=6cm,angle=0.0]{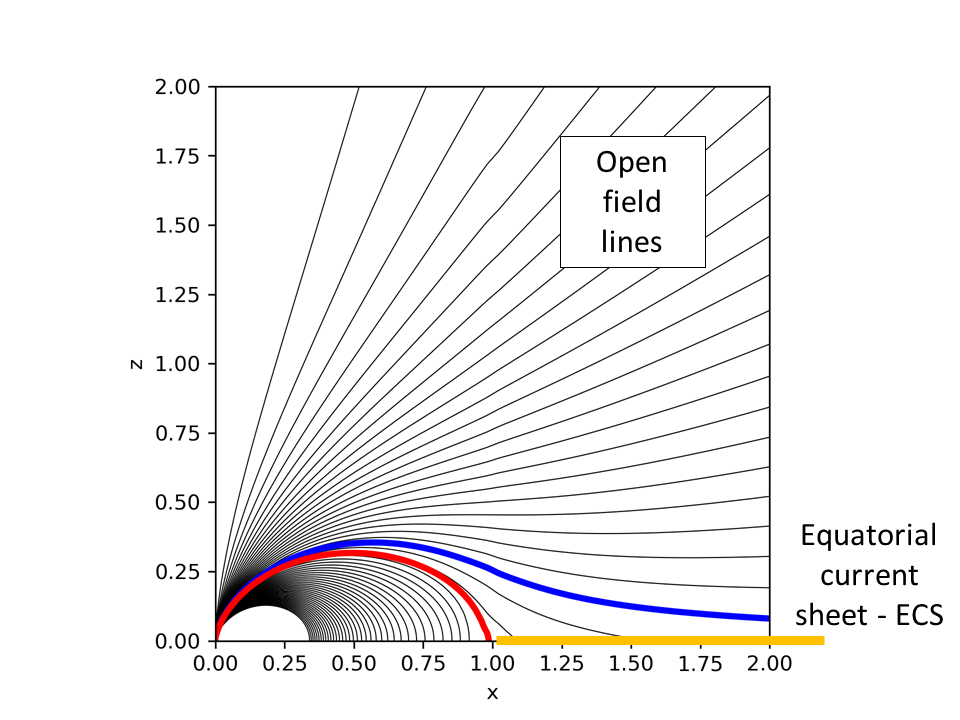}
\caption{Example of a highly dissipative magnetosphere with $\kappa=2$ (see below). Light cylinder at $x=1$. Thick red line: separatrix current sheet. Thick blue line: boundary between open field lines, and lines that enter the ECS (yellow line). Thin lines: lines of constant $\Psi$. $\Psi=0$ along the axis, and neighboring lines differ by $0.05\Psi_{{\rm open}\ {\rm dipole}}$. The thin magnetospheric zone between the red and blue lines originates in the rim of the polar cap and supplies 100\% of the electromagnetic energy that is dissipated in the ECS, and 50\% of the charges needed to support the electric charge and the electric current of the ECS (see text). Dissipation extends all the way to infinity, but most of it takes place near the light cylinder.}
\label{picture}
\end{figure}
\begin{figure}
 \centering
 \vspace{-0.3cm}
 \includegraphics[width=8cm,height=6cm,angle=0.0]{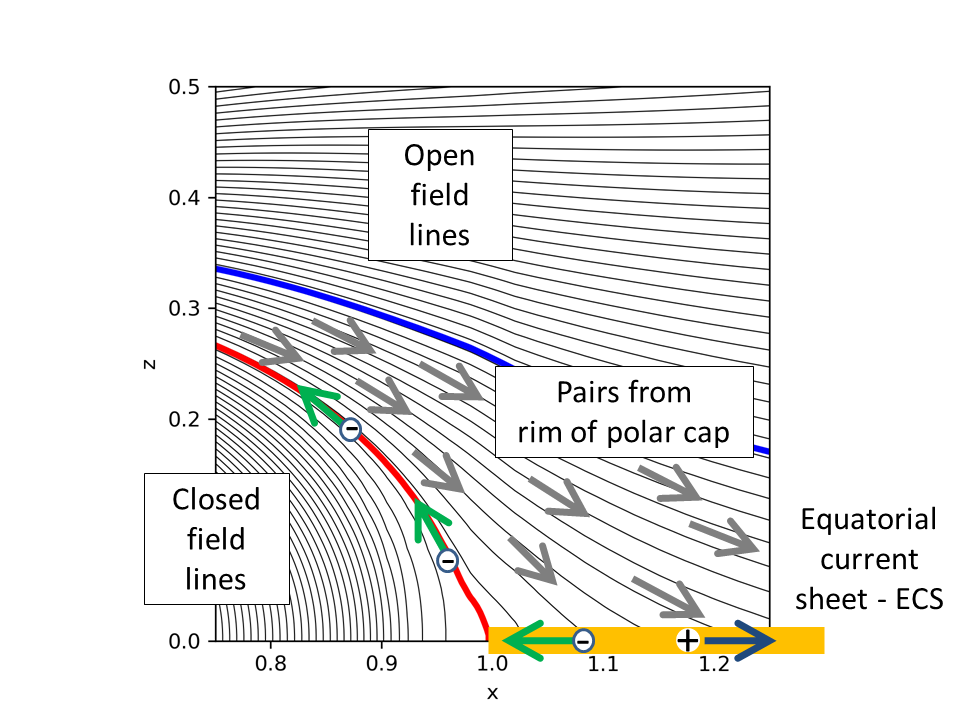}
\caption{Detail of the magnetospheric replenishment of the electric current and electric charge in the ECS near the tip of the closed-line region for the solution shown in fig.~\ref{picture}. Neighboring $\Psi$ lines differ by $0.01\Psi_{{\rm open}\ {\rm dipole}}$. Grey/green/blue arrows: pairs/electrons/positrons respectively. What is not shown here is the extra positronic electric current component that flows along the separatrix and the ECS (eq.~\ref{extraI}).}
\label{figure2}
\end{figure}

\section{Supply of pairs}

The dissipation layer extends from the tip of the closed-line region at $r\approx r_{\rm lc}$ to infinity, i.e. the dissipation layer and the ECS are one and the same. This is a natural way to connect the region of flux $\Psi_{\rm ECS}$ with the FFE solution outside (see fig.~\ref{picture}). In the limit that $\Psi_{\rm ECS}\ll \Psi_{\rm open}$, the solution must be almost indistinguishable from the dissipationless FFE solution of \citet{CKF99} with a very narrow region between the last open field lines and the separatrix and equatorial current sheets. Notice that fig.~\ref{picture} and the lower sub-figures in fig.~\ref{Multiple} below are consistent with most `ab-initio' PIC simulations  in the literature which show extended field line closure beyond the light cylinder \citep[e.g.][]{CPPS15,Ketal18}.

The ECS contains a radial electric current $I_{\rm ECS}$, has a distribution of surface electric charge density $\sigma$, and is threaded by a finite amount of magnetic flux $\Psi_{\rm ECS}$. The magnetosphere just above  the dissipation layer is an ideal force-free magnetosphere with
\begin{eqnarray}
&&E_r=-x B_z
\label{Er}\\
&&E_z=xB_r=2\pi \sigma
\label{sigma}\\
&&B_\phi=-\frac{I_{\rm ECS}}{xr_{\rm lc}c}
\end{eqnarray}
We have introduced here the notation $r/r_{\rm lc}\equiv x$.
As we discussed in Paper~I of this series, the magnetic field lines that enter the equatorial dissipation layer carry a total flux of electron-positron pairs (number of electron-positron pairs that enter the ECS per unit time and unit area) equal to $2 n_{\rm pairs}|v_z|=2 n_{\rm pairs}v_p(|B_z|/B_p)$, where $v_p, v_z$ are the poloidal and vertical component of the pair velocity, $n_{\rm pairs}$ is the number density of pairs, and $B_p$ is the poloidal magnetic field\footnote{We have assumed that there are many more pairs than primary particles in these field lines, i.e. that $\kappa\gg 1$. This allows us to ignore the electric current carried by the primary particles. In a future publication, we will generalize our analysis in the limit $1\geq \kappa\geq 0$.}. The extra factor of two is due to the two contributions from above and below the equatorial plane. These magnetic field lines originate on the polar cap, where the pairs are generated and outflow at close to the speed of light. Conservation of the pair flux implies that
\begin{eqnarray}
&&  \frac{n_{\rm pairs}v_p}{B_p}=\left. \frac{n_{\rm pairs}v_p}{B_p}\right|_*\approx\frac{\kappa\frac{\Omega B_*}{2\pi c{\rm e}}c}{B_*}=\frac{\kappa\Omega}{2\pi {\rm e}}
\end{eqnarray}
Here, $\Omega B_*/(2\pi c)\equiv \rho_{\rm GJ}$ is the Goldreich-Julian charge density at the polar caps, and ${\rm e}$ is the electron/positron charge. 
If this is the only source of charges in the dissipation layer, then the surface charge density $\sigma$ at some distance $r$ in the dissipation layer is equal to the sum of the positive surface charge density $\sigma_+$ carried by the positrons that enter the dissipation layer inside distance $r$ and move outwards towards $r$, and the negative surface charge density $\sigma_-$ carried by the electrons that enter outside distance $r$ and move inwards towards $r$. A detailed balance of the number of charge carriers that enter the ECS from above and below yields the following preliminary expression
\begin{eqnarray}
\sigma & = & \sigma_+ + \sigma_-\nonumber\\
& = &2{\rm e}\left\{\frac{\int_{r_{\rm lc}}^r2\pi r' {\rm d}r' n_{\rm pairs}|v_z|}{2\pi r v_{r+}}+\frac{-\int_r^{\infty}2\pi r' {\rm d}r' n_{\rm pairs}|v_z|}{-2\pi r v_{r-}}\right\}\nonumber\\
& \approx &\frac{2{\rm e}}{r|v_r|}\left\{\int_{r_{\rm lc}}^r r' {\rm d}r' n_{\rm pairs}|v_z|-\int_r^{\infty}r' {\rm d}r' n_{\rm pairs}|v_z|\right\}\nonumber\\
& = &\frac{2{\rm e}}{r|v_r|}\left\{2\int_{r_{\rm lc}}^r r' {\rm d}r' n_{\rm pairs}|v_z|-\int_{r_{\rm lc}}^{\infty}r' {\rm d}r' n_{\rm pairs}|v_z|\right\}
\ .
\label{balancepreliminary}
\end{eqnarray} 
Here, $v_{r+}$/$v_{r-}$ are the radial velocity of the positrons/electrons in the ECS respectively, and as we will see below, $v_{r+}\equiv |v_r|\approx -v_{r-}$ at every position along the miplane. Eq.~(\ref{balancepreliminary}) has one major flaw: as $r\rightarrow r_{\rm lc}^+$, $\sigma$ does not approach zero as it should \citep{T06}. The only way to reconcile this discrepancy, is to {\it introduce an extra outward flow of positrons through the separatrix and equatorial current sheets} equal to
\begin{eqnarray}
&&I_{{\rm ECS}\ {\rm separatrix}}=4\pi{\rm e}\int_{r_{\rm lc}}^{\infty}r' {\rm d}r' n_{\rm pairs}|v_z|\ .
\label{extraI}
\end{eqnarray}
This electric current component of the ECS may be due to electron-positron pairs that outflow along the separatrices, and when the reach the Y-point, the positrons outflow along the ECS, and the electrons flow back to the star along the separatrices. We will discuss the physical significance of this extra electric current component in a forthcoming publication. Adding the above component to eq.~(\ref{balancepreliminary}) we obtain our final expressions for the equatorial electric current density and the total equatorial electric current, namely
\begin{eqnarray}
\sigma & = & \sigma_+ + \sigma_+ 
+ \frac{I_{{\rm ECS}\ {\rm separatrix}}}{2\pi r |v_r|}\nonumber\\
& = & \frac{4{\rm e}}{r |v_r|}
\int_{r_{\rm lc}}^r r' {\rm d}r' n_{\rm pairs}|v_z|
\ ,
\label{balance}\\
I_{\rm ECS} & \approx & 2\pi r |v_r|(\sigma_+ - \sigma_-)+I_{{\rm ECS}\ {\rm separatrix}}\nonumber\\
& = & 4{\rm e}
\int_{r_{\rm lc}}^\infty 2\pi r' {\rm d}r' n_{\rm pairs}|v_z|\nonumber\\
& = & 4{\rm e}
\int_{r_{\rm lc}}^\infty 2\pi r' {\rm d}r' \left(\frac{n_{\rm pairs}v_p}{B_p}\right)|B_z|\nonumber\\
& = & 4{\rm e}\frac{\kappa\Omega}{2\pi{\rm e}}
\int_{r_{\rm lc}}^\infty 2\pi r' {\rm d}r' |B_z|\nonumber\\
& \equiv & \frac{2\kappa \Omega}{\pi}\Psi_{\rm ECS}\ .
\label{IECS}
\end{eqnarray}
Furthermore, \cite{CKF99,S06,T06} obtained numerically that
\begin{eqnarray}
&&I_{\rm ECS}\approx \frac{\Omega \Psi_{{\rm open}\ {\rm dipole}}}{2\pi}=\frac{1}{2}\Omega B_* r_{{\rm pc}\ {\rm dipole}}^2\ .
\label{CKFS06}
\end{eqnarray}
This very interesting numerical result has never before been pointed out in the literature\footnote{As is well know since \cite{CKF99}, the electric current distribution along open magnetic field lines has a maximum value near the maximum electric current of a split monopole magnetic field configuration with the same amount of open magnetic flux $\Psi_{\rm open}$, namely 
$\Omega \Psi_{\rm open}/(2\pi)$. Beyond that maximum, the magnetosphere contains a region of return electric current near the equator. We now point out for the first time that, the amount of return electric current is such that the remaining return current that flows along the equatorial current sheet is equal to $\Omega \Psi_{{\rm open}\ {\rm dipole}}/(2\pi)$, and {\it not} $\Omega \Psi_{\rm open}/(2\pi)$ as would be naively expected from the analogy with a split monopole configuration.}. Reversing eq.~(\ref{IECS}) and using eq.~(\ref{CKFS06}) above, we obtain the amount of magnetic flux $\Psi_{\rm ECS}$ along the rim of the polar cap that contains the electric charges needed in the equatorial current sheet, namely
\begin{eqnarray}
&&\Psi_{\rm ECS}=\frac{2\delta}{r_{\rm pc}}\Psi_{\rm open}=\frac{\pi I_{\rm ECS}}{2\kappa\Omega}\approx \frac{\Psi_{{\rm open}\ {\rm dipole}}}{4\kappa}\approx\frac{\Psi_{\rm open}}{5\kappa}\ .
\end{eqnarray}
This relation allows us to obtain the thickness $\delta$ of the rim of the polar cap, namely
\begin{eqnarray}
&&\delta\approx \frac{r_{\rm pc}}{10\kappa}\ .
\end{eqnarray}
Note that the above detailed considerations yielded a correction in the expression for $\delta$ with respect to the one in Paper~I (eq.~9). 

\section{Particle orbits in the ECS}

\begin{figure}
 \centering
 \includegraphics[width=12cm, angle=270.0]{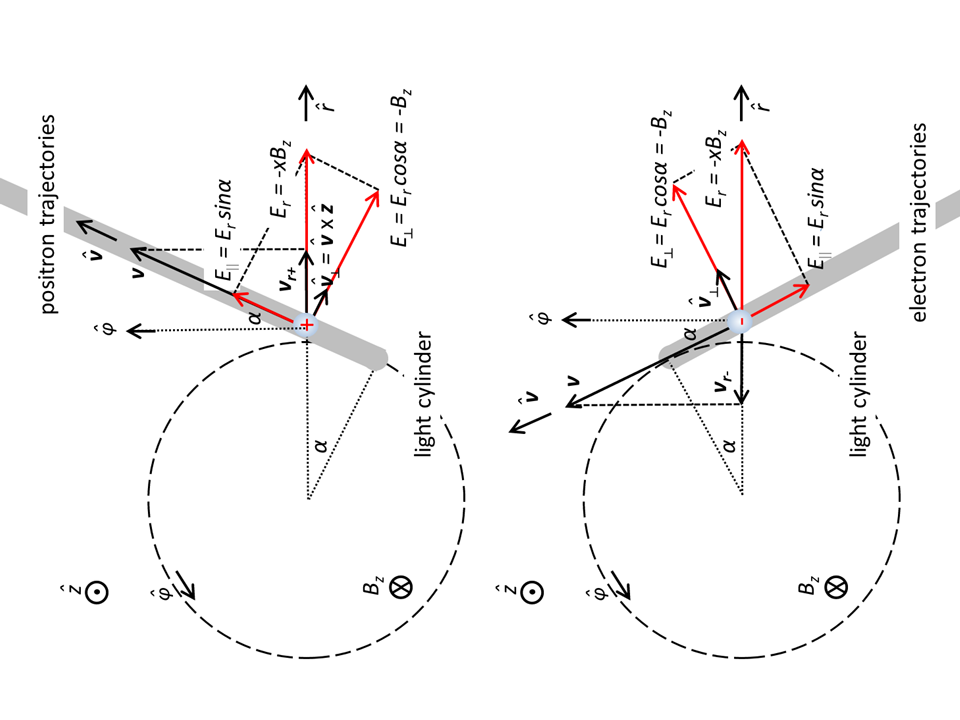}
\caption{Positron (top) and electron (bottom) trajectories (thick grey lines) in the ECS along the equatorial plane $z=0$ seen from above. Vector notation as in text.  Black arrows: velocity components. Red arrows: electric field components. $|v|\approx c$. Both trajectories are straight lines tangential to the light cylinder (dashed circle). Positrons that enter the ECS move along $\hat{\phi}$ and outwards. Electrons that enter the ECS move along $\hat{\phi}$ and inwards. Electron trajectories end on the light cylinder.}
\label{detail}
\end{figure}

\begin{figure*}
 \vspace{-0cm}
 \includegraphics[width=24cm, angle=270]{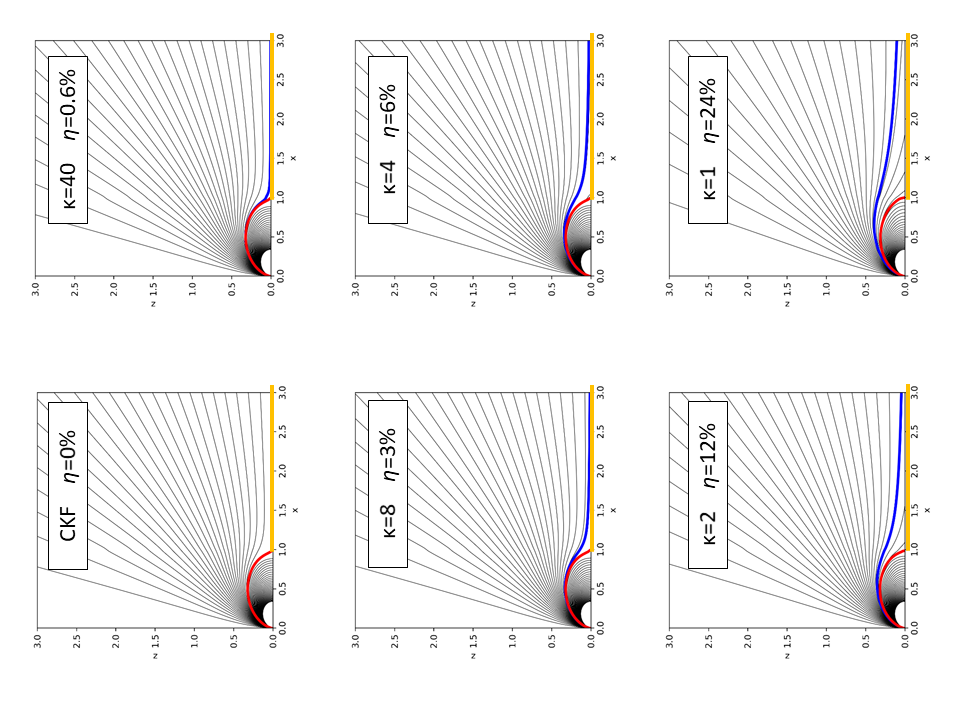}
 
 
 
 
 
 
 
 
 

 
 \vspace{-0.5cm}
 
\caption{Magnetospheric structure for various values of $\kappa\gsim 1$. $\eta\equiv \dot{E}_{\rm ECS}/\dot{E}$ is the corresponding dissipation efficiency. Red line: separatrix. Lines as in figure~1. Clockwise from top left: $\kappa=\infty, 40, 4, 1, 2, 8$. }
\label{Multiple}
\end{figure*}

Let us now consider the motion of electrons and positrons at the mid-plane of the dissipation layer beyond the light cylinder. Electrons and positrons do not just move radially. They move very close to the speed of light, but they are also deflected in the azimuthal direction together with the overall pulsar rotation. At the mid-plane, $B_\phi=0$ and $E=E_r=x|B_z|>|B_z|=B$. The total electromagnetic force acting on the positrons in the mid-plane is equal to
\begin{eqnarray}
&& {\rm e}(E_r \hat{r}+|v|B_z \hat{\bf v}\times \hat{z}/c)\approx {\rm e}(E_r \hat{r}+B_z \hat{\bf v}\times \hat{z})
\label{EM1}
\end{eqnarray}
(vectors with hats denote unit vectors along them). For an extremely relativistic particle with $|v|\approx c$, the total electromagnetic force must be equal to
\begin{eqnarray}
&& m_{\rm e} \frac{{\rm d}(\Gamma {\bf v})}{{\rm d}t}
= m_{\rm e} c\frac{{\rm d}\Gamma}{{\rm d}t}\ \hat{\bf v}+m_{\rm e}\frac{\Gamma c^2}{R_{\rm c}}\ \hat{\bf v}_\perp\ .
\label{force}
\end{eqnarray}
The second term in the above expression is the centrifugal force. Here, $R_{\rm c}$ is the radius of curvature of the particle orbit in the equatorial plane, and $\hat{\bf v}_\perp\equiv \hat{\bf v}\times \hat{z}$ is the unit vector away from the center of the instantaneous circular orbit. We will henceforth make the approximation that the instantaneous radius of curvature is so large that the centrifugal force term is much smaller than the parallel acceleration term. Decomposing eq.~(\ref{EM1}) along $\hat{\bf v}$ and $\hat{\bf v}_\perp$ we obtain 
\begin{eqnarray}
{\rm e}(E_r \hat{r}+B_z \hat{\bf v}\times \hat{z}) & = & {\rm e}E_r (\hat{r}\cdot \hat{\bf v})\hat{\bf v}+{\rm e}(E_r (\hat{r}\cdot \hat{\bf v}_\perp)+B_z )\hat{\bf v}_\perp\nonumber\\
& = & {\rm e}E_{||} \hat{\bf v}+{\rm e}(E_\perp+B_z )\hat{\bf v}_\perp
\ .
\label{EM2}
\end{eqnarray}
The expressions in eqs.~(\ref{EM2}) and (\ref{force}) must be equal to each other, and therefore, the term along $\hat{\bf v}_\perp$ must almost vanish. Thus, $E_\perp+B_z\approx 0$, and since $E_\perp=E_r \cos\alpha =-xB_z \cos\alpha$, 
\begin{eqnarray}
&& \cos\alpha=\frac{1}{x}\ .
\end{eqnarray}
Here, $\alpha$ is the angle between the azimuthal direction $\hat{\phi}$ and the direction of particle motion $\hat{\bf v}$. 
We remind the reader that, beyond the light cylinder, $E>B$ in the equatorial plane. Similar considerations apply to the electrons in the ECS. From the above, one can easily show that
\begin{eqnarray}
|v_r|&\equiv & |v|\sin\alpha \approx \frac{\sqrt{x^2-1}}{x}c\ ,
\label{vr}\\
v_\phi &\equiv & |v|\cos\alpha \approx \frac{1}{x}c\ .
\label{vphi}
\end{eqnarray}
\footnote{Note added in proof: these are the same as the components of the so-called `Aristotelian' speed of light velocities for electrons and positrons in the ECS postulated by \cite{G12}.}With the above two equations we reach the following unexpected result: after the electrons and positrons enter the ECS, they follow straight lines that are tangential to the light cylinder! The positrons travel outwards whereas the electrons travel inwards. Both travel along the direction of pulsar rotation (see fig.~\ref{detail} for details).  The closer we are to the light cylinder, the more tangential the orbits, and the further away, the more radial they are. Straight lines have an infinite radius of curvature, and therefore, eqs.~(\ref{vr}) and (\ref{vphi}) are exact. It would be nice to check whether particle trajectories in the ECS are also along straight lines in PIC numerical simulations \citep[e.g.][]{CPS16,Ketal18}.

The raison d'$\hat{\rm e}$tre of the above discussion is that we prefer to avoid the complex integration of the Speiser-like orbits that the particles follow when they enter the ECS \citep[][see paper~II]{S65}. After all, as the particles gain energy, they are confined more and more towards the mid-plane of the ECS where $B_\phi=0$. We thus ignored the meandering motion due to the azimuthal component of the magnetic field $B_\phi$ in a guiding center-type approximation. In a forthcoming publication, when we will consider the effect of radiation reaction in the particles' motion, we will need to evaluate the radius of curvature of the meandering particle trajectory. 

Putting everything together and differentiating eq.~(\ref{balance}) we obtain
\begin{eqnarray}
\frac{{\rm d}}{{\rm d}r}(r|v_r|\sigma) & = &\nonumber\\
\frac{{\rm d}}{{\rm d}x}(\sigma \sqrt{x^2-1}c) & = 
& 4{\rm e}r n_{\rm pairs}|v_z|
= 4{\rm e}r n_{\rm pairs}v_p(|B_z|/B_p)\nonumber\\
& = & \frac{2\kappa\Omega r}{\pi}|B_z|\ .
\label{diff}
\end{eqnarray}
Solving for the distribution of $B_z$ along the dissipation layer, and remembering that $\sigma=E_z/(2\pi)=xB_r/(2\pi)$ yields
\begin{eqnarray}
&&B_z=-\frac{1}{4\kappa x}
\frac{\rm d}{{\rm d}x}(x\sqrt{x^2-1}B_r)\ .
\label{Bz}
\end{eqnarray}
Notice that $B_z$ is negative. 
The latter simple result is the basis of the hybrid approach proposed below that yields the ideal force-free magnetosphere with a realistic dissipative equatorial boundary condition.

\section{Hybrid numerical method}

The new element of the present work is that the ECS and the dissipation layer are one and the same (or in other words that the magnetic flux $\Psi_{\rm ECS}$ from the rim of the polar cap is distributed all along the ECS). The realization that the ECS is not dissipationless modifies the global solution in a subtle way. We propose the following iterative numerical approach that allows us to obtain a self-consistent global solution that is ideal force-free everywhere except in the ECS:
\begin{enumerate}
\item We use the solver introduced in \citet{CKF99} to solve the pulsar equation. This allows us to obtain the unique axisymmetric ideal force-free magnetospheric solution that crosses the light cylinder smoothly for a particular equatorial boundary condition beyond the light cylinder \citep[see also e.g.][]{C07a,C07b}.
\item We obtain first the dissipationless solution of \citet{CKF99} (the so-called CKF solution) by setting $\Psi=\Psi_{\rm open}$ along the equator beyond the light cylinder, and iteratively adjusting the value of $\Psi_{\rm open}$. This solution contains a dissipationless equatorial return current sheet connected to two separatrix return current sheets at the Y-point that develops at the tip of the corotating closed-line region.
\item From the solution, we obtain the distribution of $B_r$ just above the ECS. Then, according to eq.~(\ref{Bz}),
\begin{eqnarray}
\Psi(r\ge r_{\rm lc})&=&\Psi(r_{\rm lc})+\int_{r_{\rm lc}}^{r}2\pi  B_z\ r'{\rm d}r'\nonumber \\
&=&\Psi_{\rm open}-\frac{\pi r_{\rm lc}^2}{2\kappa}
x\sqrt{x^2-1}B_r\ .
\label{PsiECS}
\end{eqnarray}
\item Given this new Dirichlet-type boundary condition along the ECS, we solve again the pulsar equation above the ECS. This yields a new $B_r$ distribution.
\item We repeat the above steps (iii) and (iv) till the solution relaxes to a steady-state configuration in which both the electric current and the electric charge of the ECS are accounted-for self-consistently, and  eq.~(\ref{diff}) is satisfied everywhere along the ECS. 
\end{enumerate}

\begin{figure}
 \centering
 \vspace{-0.3cm}
 \includegraphics[width=9cm,angle=0.0]{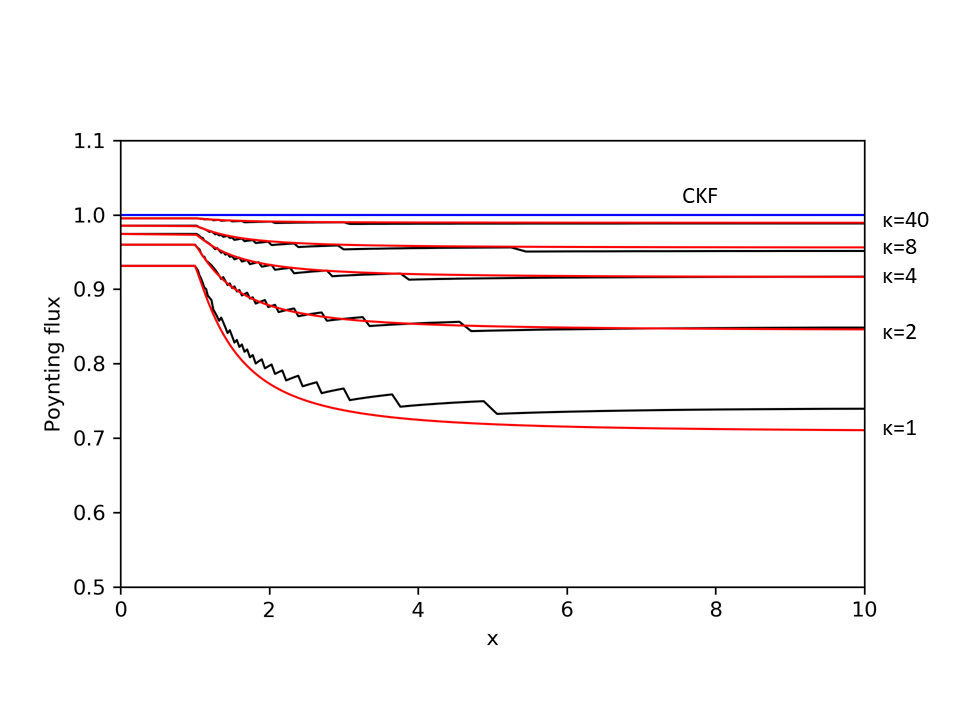}
 \vspace{-0.6cm}
\caption{Outgoing Poynting flux integrated over a sphere of radius $x$ as a function of radius for various values of $\kappa$ (black lines) and corresponding analytical fits according to eq.~(\ref{Edotan}) (red lines). Energy flux  normalized to the spin-down power of an aligned pulsar without dissipation (CKF solution; blue line). Most dissipation takes place within about two light cylinder radii from the light cylinder, and exceeds $20\%$ of $\dot{E}$ only for old pulsars with very low pair-formation multiplicity.}
\label{Poynting}
\end{figure}

We implemented the above numerical method and obtained the global magnetospheric structure of an aligned pulsar rotator for various values of the pair formation multiplicity parameter $\kappa\geq 1$ (fig.~\ref{Multiple}). Each iteration runs on a $200\times 200$ spatial numerical grid and takes about one hour to converge.
The stellar dipole boundary condition is imposed in the central circle of radius $0.1 r_{\rm lc}$. The separatrix return current sheet has a width of about $0.05 \Psi_{\rm open}$ inside the red lines of figs.~\ref{picture}, \ref{figure2} and \ref{Multiple}. For $\kappa\gsim 40$, the solution is almost indistinguishable from the ideal solution of \cite{CKF99}. In that case, the calculation of dissipation, particle acceleration and high energy radiation can only be based on analytical approximations of the equatorial electric and magnetic fields (see eqs.~\ref{Bran}-\ref{Bphian} below). Notice that our analysis is valid for $\kappa\gsim 2$ since below that value, our approximation that $\Psi_{\rm ECS}=(5\kappa)^{-1}\Psi_{\rm open}\ll \Psi_{\rm open}$ breaks down. We also calculated the outgoing Poynting flux integrated over a sphere of radius $x$ as a function of radius for various values of $\kappa$ (fig.~\ref{Poynting}). Most dissipation takes place within about two light cylinder radii from the light cylinder, and exceeds a few tens of percent of $\dot{E}$ only for old pulsars with extremely low pair-formation multiplicity.

Notice the similarity between case $\kappa=1$ in fig.~\ref{Multiple} and case `$f_{inj}=1$' in figure~3 of \cite{CPPS15}, as well as between fig.~\ref{Poynting} and figure~6 of that paper. This similarity is by itself very interesting. It implies that, global PIC simulations with the lowest possible (numerically) amount of dissipation shown in the literature \citep[e.g.][]{CPS16,Ketal18} are very similar to our dissipative solutions which describe pulsars with very low pair formation multiplicities $\kappa\approx 1$, {\it not} young pulsars with $\kappa\gg 1$. This confirms our concern that ab-initio numerical simulations are presently inadequate to study the physical electromagnetic energy dissipation in the pulsar magnetosphere. Our hybrid method, however, allows us to have better control over the numerical dissipation since the bulk of the magnetosphere is by construction ideal, and dissipation is restricted to the ECS. This is why we are able to run simulations with extremely low dissipation and very high $\kappa$ values.

\section{Useful approximations}

In young pulsars with high pair-formation multiplicity $\kappa\gg 1$, the distribution of $B_r$ just above the ECS that we obtained numerically with the above procedure may be approximated by the expression
\begin{eqnarray}
B_r&\approx& 
\frac{1}{x^2}\left(1-\frac{1}{x^2}\right)^{0.7}
B_{{\rm lc}\ {\rm dipole}}
\ .
\label{Bran}
\end{eqnarray}
Here, $B_{{\rm lc}\ {\rm dipole}}\equiv B_* r_*^3/(2r_{\rm lc}^3)$ is the equatorial value of the vacuum dipole magnetic field at the light cylinder.
Therefore, according to eqs.~(\ref{Er}) and (\ref{Bz})
\begin{eqnarray}
B_z&\approx & 
-\frac{3}{5\kappa x^4}\left(1-\frac{1}{x^2}\right)^{0.2}
B_{{\rm lc}\ {\rm dipole}}
\ ,
\label{Bzan}\\
E_r 
&\approx & \ \ \ \!\frac{3}{5\kappa x^3}\left(1-\frac{1}{x^2}\right)^{0.2}
B_{{\rm lc}\ {\rm dipole}}\ .
\label{Eran}
\end{eqnarray}
Notice the very sharp decrease of $B_z$ and $E_r$ with distance.
Finally, let us also introduce
\begin{eqnarray}
B_\phi&=&-\frac{I_{\rm ECS}}{xr_{\rm lc}c}
= -\frac{\Omega r_{{\rm pc}\ {\rm dipole}}^2 B_*}{2x r_{\rm lc} c}
= -\frac{B_{{\rm lc}\ {\rm dipole}}}{x}
\ .
\label{Bphian}
\end{eqnarray}
This is a nice simple result that derives from eq.~(\ref{IECS}). 
We can now obtain analytically the distribution of electromagnetic (Poynting) flux that enters the ECS, namely
\begin{eqnarray}
\dot{E}_{\rm ECS}&=&2\int_{x=1}^x 2\pi r_{\rm lc}^2 \frac{c}{4\pi}E_r |B_\phi|\ x{\rm d}x\nonumber\\
&=& \frac{I_{\rm ECS}}{2\pi r_{\rm lc}} \int_{x=1}^x 2\pi r_{\rm lc}^2 |B_z|\ x{\rm d}x\nonumber\\
&\approx& \frac{I_{\rm ECS}\Psi_{\rm ECS}}{2\pi r_{\rm lc}} \left(1-\frac{1}{x^2}\right)^{1.2}\nonumber\\
&\approx & \frac{6}{25\kappa}\dot{E}\left(1-\frac{1}{x^2}\right)^{1.2}\ .
\label{EdotECS}
\end{eqnarray}
The factor of two in eq.~(\ref{EdotECS}) takes into account the fact that both hemispheres emit Poynting flux. Here,
$\dot{E}\approx (2/3)\Omega^2 (\Psi_{\rm open}/2 \pi)^2/c$ is the total electromagnetic spindown energy loss rate \citep{CS06}.
Equivalently, the outgoing Poynting flux integrated over a sphere of radius $r$ is equal to
\begin{eqnarray}
\dot{E}_{\rm Poynting}(x)&=&\dot{E}-\dot{E}_{\rm ECS}(x)\nonumber\\
&\approx &
\begin{cases}
\ \ \dot{E}\left(1-\frac{6}{25\kappa}\left(1-\frac{1}{x^2}\right)^{1.2}\right)        &\text{if $x \geq 1$}\\
\ \ \dot{E}         &\text{otherwise}
\end{cases}
\label{Edotan}
\end{eqnarray}
As we can see in fig.~\ref{Poynting}, the fits are almost perfect for $\kappa\gsim 2$, and break down for $\kappa\leq 1$. Most of the particle acceleration and consequent radiation in the ECS take place very close to the light cylinder, hence the justification of the term `ring-of-fire' introduced in Paper~II.

Up to now, we have assumed that the pair formation multiplicity $\kappa$ is very high. However, in order to attain observed dissipation efficiencies on the order of 1 to $10\%$, we need $\kappa$ values on the order of 20 to 2. We suspect that these are not typical values for the bulk of the polar cap, and that $\kappa\rightarrow 0$ as we approach the edge of the polar cap along the separatrix between field lines that close inside and outside the light cylinder. This idea certainly needs further investigation.

\section{Conclusion}

In this series of three papers, we associate the magnetospheric dissipation with the `struggling' of the magnetosphere to supply the electric charges required to support the electric charge and the electric current of the equatorial current sheet (ECS). During our self-consistent investigation we discovered that the supply of pairs from the rims of the polar caps is not sufficient. The ECS requires an extra amount of positronic electric current that originates in the stellar surface and flows outwards along the separatrices. We will discuss the physical significance of this extra positronic electric current in a forthcoming publication.

The hybrid numerical method presented in this work allows us to study the magnetospheric dissipation in a realistic pulsar magnetosphere at a level never before being possible with standard numerical simulations (field calculations and ab-initio PIC calculations). We have obtained analytical expressions for the distribution of dissipation along the ECS as a function of the pair-formation multiplicity $\kappa$. As shown also in several previous works, magnetospheric dissipation indeed takes place within a couple of light cylinder radii beyond the tip of the closed-line region at the light cylinder, hence the name `ring-of-fire' introduced in Paper~II of this series.

The analytical expressions for $B_z$ and $E_r$ in the ECS that we derived above allow us to calculate directly, not only the distribution of dissipated electromagnetic energy, but also the detailed outward acceleration of the positrons and the inward acceleration of the electrons in the ECS in the presence of radiation reaction. For the particular straight line motion along the ECS discussed in \S~3 above, if we consider the radius of curvature $R_{\rm c}$ of the meandering motion above and below the equator, the force balance equation along the instantaneous direction of motion in the presence of radiation reaction (eq.~14, Paper~I)
becomes
\begin{eqnarray}
&&\frac{{\rm d}\Gamma}{{\rm d}x} 
= \frac{{\rm e}B_{\rm lc}r_{\rm lc}}{m_{\rm e}c^2}\left\{ \frac{3}{5\kappa x^{3}}\left(1-\frac{1}{x^2}\right)^{0.2}-\frac{\Gamma^4/\Gamma_{\rm rrl}^4}{(R_{\rm c}/r_{\rm lc})^4}\left(1-\frac{1}{x^2}\right)^{-0.5}\right\}\nonumber \\
\label{dGammadx}
\end{eqnarray}
Here, $\Gamma_{\rm rrl}\equiv (3 r_{\rm lc}^2 B_{\rm lc}/2{\rm e})^{1/4}=4\times 10^7(B_*/10^{13}{\rm G})^{1/4}(P/1\ {\rm s})^{-1/4}$, 
and $P$ is the pulsar period. 
The integration of eq.~(\ref{dGammadx}) will yield the spectrum of the emitted $\gamma$-ray radiation, and will be performed in a forthcoming publication.


\section*{Acknowledgements}

P.S. would like to acknowledge support from PHAROS COST Action CA16214 for a Short-Term Scientific Mission at the Observatoire Astronomique de Strasbourg in July-August 2019.


\bibliographystyle{mn2e}

\end{document}